# Ionic Liquid Biospheres

**Sara Seager** [1,2,3,4], **William Bains** [5], **Iaroslav Iakubivskyi** [1], **Rachana Agrawal** [6], **John Jenkins** [5], **Pranav Shinde** [5] and **Janusz J. Petkowski** [7,8,*]

[1] Department of Earth, Atmospheric and Planetary Sciences, Massachusetts Institute of Technology, 77 Massachusetts Avenue, Cambridge, MA 02139, USA
[2] Department of Physics, Massachusetts Institute of Technology, 77 Massachusetts Avenue, Cambridge, MA 02139, USA
[3] Department of Aeronautical and Astronautical Engineering, Massachusetts Institute of Technology, 77 Massachusetts Avenue, Cambridge, MA 02139, USA
[4] Kavli Institute for Astrophysics and Space Research, Massachusetts Institute of Technology, 77 Massachusetts Avenue, Cambridge, MA 02139, USA
[5] School of Physics and Astronomy, Cardiff University, 4 The Parade, Cardiff CF24 3AA, UK
[6] Department of Aerospace Engineering, Indian Institute of Technology Kanpur, Kalyanpur, Kanpur 208016, UP, India
[7] Faculty of Environmental Engineering, Wroclaw University of Science and Technology, 50-370 Wroclaw, Poland
[8] JJ Scientific, 02-792 Warsaw, Poland
* Correspondence: janusz.petkowski@pwr.edu.pl

**Abstract**

Liquid is a fundamental requirement for life as we understand it, but whether that liquid has to be water is not known. We propose the hypothesis that ionic liquids (ILs) and deep eutectic solvents (DES) constitute a class of non-aqueous planetary liquids capable of persisting on a wide range of bodies where stable liquid water cannot exist. This hypothesis is motivated by key physical properties of ILs and DES. Many exhibit vapor pressures orders of magnitude lower than that of water and remain liquid across exceptionally wide temperature ranges, from cryogenic to well above terrestrial temperatures. These properties permit stable liquids to exist where liquid water would rapidly evaporate or freeze and outside of bulk phases as persistent microscale reservoirs—such as thin films and pore-filling droplets. In other words, ILs and DES can persist in environments without requiring oceans, thick atmospheres, or narrowly regulated climate conditions. We further hypothesize that ILs and DES could act as solvents for non-Earth-like life, based on their polar nature and the demonstrated stability and functionality of proteins and other biomolecules in ionic liquids. More speculatively, our hypothesis extends to the idea that ILs and DES could enable prebiotic chemistry by providing long-lived, protective liquid environments for complex organic molecules on bodies such as comets and asteroids, where liquid water is absent. Additionally, based on the occurrence of DES-like mixtures as protective intracellular liquids in desiccation-tolerant plants, we propose that ILs and DES might be solvents that life elsewhere purposefully evolves. We review protein and other biomolecule studies in ILs and DES and outline planetary environments in which ILs and DES might occur by discussing available anions and cations. We present strategies to advance the IL/DES solvent hypothesis using laboratory studies, computational chemistry, planetary missions, analysis of existing spectroscopic datasets, and modeling of liquid microniches and chemical survival on small bodies.

## 1. Motivation for Considering Ionic Liquids and Deep Eutectic Solvents as Planetary Liquids

Liquid is very likely a fundamental requirement for life as we understand it, but whether that liquid has to be water is not known [1–3]. Many planetary bodies are hostile to long-lived surface water: atmospheres are thin or absent, temperatures swing widely, and liquid reservoirs are transient or unstable. As a result, large fractions of the Solar System and the exoplanet population are excluded from classical habitability considerations, simply because liquid water cannot persist in bulk. This challenge creates a strong motivation to consider alternative liquids whose physical properties allow them to remain stable where liquid water cannot.

Ionic liquids (ILs) and deep eutectic solvents (DES) are compelling as alternative planetary liquids because of their physical properties. Many exhibit extremely low vapor pressures compared to water and remain liquid across wide temperature ranges, from cryogenic to well above terrestrial conditions (e.g., [4,5]). For example, ionic liquid tributylammonium-1,3-di-(trifluoromethyl)-pentane-2,4-diketonate [NBu₃H][HFAC], has one of the lowest known melting points for an IL, melting at −93 °C [6] (Figure 1; Table 1). These properties allow ILs and DES to exist outside of bulk phases, as stable films, veins, and microdroplets under conditions that cannot support liquid water. In effect, liquid-phase chemistry can persist without requiring oceans, thick atmospheres, or tightly regulated climate cycles, opening vast new classes of planetary environments—from airless bodies to hot terrestrial exoplanets and cometary interiors.

**Table 1.** Physical properties of selected ILs and DES. Individual compounds differ significantly from each other with respect to melting points, densities, viscosities, or decomposition temperatures. The IL and DES unifying characteristics are generally very low vapor pressure and high viscosity as compared to liquid water, with ILs generally having significantly lower vapor pressures than DES. The (—) symbol indicates no available data.

| Selected ILs and DES | Melting Point (°C) | Density (g/cm³ at 20–25 °C) | Viscosity (mPa·s ~25 °C) | Decomp. Temp. (°C) | Vapor Pressure (Pa at Temp. °C) | Ref. |
|---|---|---|---|---|---|---|
| [EMIM][BF$_4$] | 15 | 1.28 | 33.8 | 200–300 | — | [7] |
| [BMIM][PF$_6$] | −8 | 1.37 | 310 | >200 | $1 \times 10^{-10}$ at 25 °C | [7,8] |
| [EMIM][TFSI] | −17 | 1.53 | 39.4 | 250 | $8.9 \times 10^{-12}$ at 25 °C | [7,9] |
| [BMIM][DCA] | −6 | 1.06 | 28 | — | $1.5 \times 10^{-13}$ at 25 °C | [9] |
| [EA][NO$_3$] | 12 | 1.26 | 28 | 250 | — | [10] |
| [NBu$_3$H][HFAC] | −93 | 0.68 | 16.80 | — | — | [6] |
| sorbitol/urea/NH$_4$Cl (7:2:1) | 67 | — | — | 220 | 126.1 at 70 °C | [11] |
| choline chloride/urea (1:2) | 12 | 1.24 | 632 | >200 | 2.9 at 120 °C | [12,13] |
| K$_2$CO$_3$/glycerol (1:4) | −40 | 1.48 | 28,000 | — | — | [14] |
| **Water** | 0 | 0.99 | 0.89 | >2000 | 3169 | [15] |

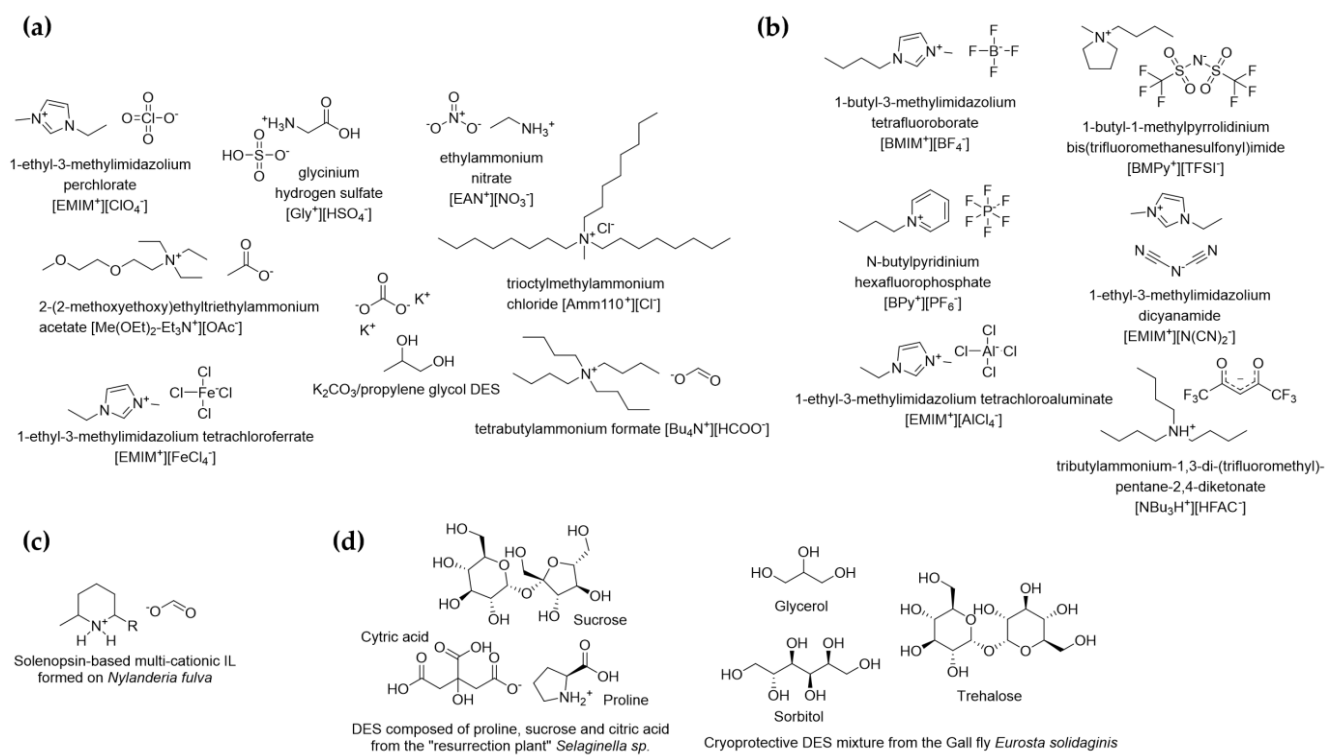

**Figure 1.** Chemical structures of selected ILs and DES. (**a**) Anions common to planetary environments could lead to abiotic formation of ILs and DES, including, e.g., NO$_3^-$, HSO$_4^-$, Cl$^-$, ClO$_4^-$, HCOO$^-$, CH$_3$COO$^-$, FeCl$_4^-$, and CO$_3^{2-}$. In principle, any positively charged

nitrogen-containing organic molecule could act as the cation in a "planetary" ionic liquid pair; (**b**) selected industrial synthetic ionic liquids not expected to be present in planetary settings; (**c**) the known natural ionic liquid formed between two ant species as a venom counteragent; (**d**) examples of selected natural deep eutectic solvents (NADES) mixtures made by desert plants and fly larvae. Note that NADES mixtures can be very diverse in their chemical composition and molar ratios of individual components, differing significantly even between closely related species. References (**a**): [16–23] (**b**): [6,24–28] (**c**): [29] (**d**): [30,31].

We propose the hypothesis that ILs and DES constitute a class of naturally occurring planetary liquids, providing habitable microscale environments even on bodies where liquid water is absent [17]. We further speculate that ILs and DES could act as solvents for non-Earth-like life based on their polar nature, making them suitable for selective dissolution of organic molecules. Additionally, based on the occurrence of DES-like mixtures as protective intracellular liquids in desiccation-tolerant plants, we hypothesize that ILs and DES might be solvents that life elsewhere purposefully evolves.

From a planetary and astrobiology perspective, we treat ionic liquids (ILs) and deep eutectic solvents (DES) together. Their unifying feature for our purposes is that they are liquids without bulk water and in conditions where liquid water may not be stable. For questions of planetary habitability, the useful distinctions between ILs and DES are therefore secondary—mainly the extent of vapor pressure suppression and whether their components could plausibly form abiotically on a planetary body. Chemically, ILs and DES can be very distinct. Ionic liquids are defined as salts that are liquid below 100 °C; they are strongly bonded together by ionic bonds [32]. DES are compound mixtures that have a much lower melting point than individual components and are held together by a variety of electrostatic forces, including hydrogen bonding, van der Waals interactions, dipole–dipole interactions, and sometimes ionic bonding [13]. For a discussion of differences between ILs and DES, see [33].

Although thousands of ionic liquids and DES have been developed for industrial applications, most rely on complex synthetic organic ions or additives that are unlikely to arise under natural planetary conditions (e.g., [13,34,35]). However, recent work [17,36] has highlighted a subset of ionic liquids and closely related mixtures that can form from cosmically abundant species, including simple inorganic ions, small organics, and strong acids. While many ILs and DES are known to be highly hygroscopic (e.g., [37]), our hypothesis focuses on bodies without liquid water or appreciable water vapor, where the ILs and DES would persist in largely anhydrous form.

We explore our proposed hypothesis by first reviewing existing evidence relevant to the compatibility of ionic liquids and deep eutectic solvents with biomolecules, including protein stability in ILs and the occurrence of DES-like mixtures as functional intracellular liquids in desiccation-tolerant plants (Section 2). We then examine implications for planetary exploration by identifying cosmically available anions and cations capable of forming simple planetary ILs and DES, and by surveying planetary environments in which these liquids could plausibly occur, including Mars, Venus' sub-cloud layers, asteroids, comets, icy bodies, and water-depleted exoplanets (Section 3). We present speculative hypotheses for the roles of ILs and DES in prebiotic chemistry and solvent evolution (Section 4). Finally, we present strategies to substantiate the hypothesis, including targeted laboratory experiments, computational chemistry, reanalysis of existing spacecraft and astronomical datasets, and the development of mission concepts and models to test the formation, stability, detectability, and chemical persistence of IL- and DES-based liquid microniches in planetary settings (Section 5).

## 2. Review of ILs and DES as Biologically Relevant Liquids

We summarize current knowledge in support of our hypothesis, without intending to be exhaustive. We begin by briefly reviewing the chemical compatibility of proteins (Section 2.1) and other biomolecules (Section 2.2) with ionic liquids, followed by a summary of DES-like mixtures as functional intracellular liquids in desiccation-tolerant plants (Section 2.3).

*2.1. Compatibility of Proteins with Ionic Liquids*

A minimal requirement for any alternative solvent for life is compatibility with complex organic molecules, particularly some of the polymers and small organic molecule cofactors central to terrestrial biochemistry. Demonstrating such compatibility is therefore essential for evaluating the hypothesis that ionic liquids could function as life-supporting media. Ionic liquids satisfy this criterion to a remarkable degree, despite having very different physics and chemistry from water, including a strongly ionic character.

We carried out a literature survey of protein stability and function in ionic liquids and identified 109 distinct ionic liquids, comprising 53 unique cations and 30 unique anions, in which the structural stability or activity of 52 different proteins has been experimentally tested (original data and references compiled in the Supplementary Materials). To isolate the protein behavior in genuinely non-aqueous environments, we include only experiments that have been performed in ionic liquids containing ≤5% water; 85% of the studies fall in this category. Remarkably, 71% of tested proteins retain their native folded three-dimensional structure, 70% are soluble, and 65% of enzymes retain measurable catalytic activity when assayed. The majority of protein stability and function tests were performed at room temperature. Remarkably, in a few cases, e.g., for the enzyme cellulase, the protein catalytic activity in ILs was retained at 65 °C. Cellulase itself is stable and remains folded in ILs at temperatures as high as 115 °C [38].

The takeaway is that some ionic liquids are readily compatible with some Earth-life biochemical machinery, even if tested proteins do not have any specific, pre-developed, evolutionary adaptations toward functioning in such varied, non-aqueous chemical environments. We emphasize that the examples of compatibility of proteins with ionic liquid are just illustrative and do not mean that ionic liquid biospheres have to rely on Earth-life proteins to function. If life were to exist in ionic liquids, it would not have to employ the same molecular machinery as terrestrial life, particularly given the much broader temperature ranges over which many ionic liquids remain stable.

*2.2. Compatibility of Non-Protein Biomolecules with Ionic Liquids*

Beyond proteins, a large number of biomolecules are stable and soluble in a variety of ionic liquids and DES [39–41]. Both low-molecular-weight sugars and complex carbohydrates are generally soluble and stable in ionic liquids [19,38,42–50]. DNA is also stable in pure ionic liquids, and in some cases, ILs stabilize the DNA double helix structure (reviewed in [51]), protect it from degradation [52,53], or even form solutions that contain duplex DNA as the anion in the ionic liquid anion-cation pair [54].

ILs are rarely seen in terrestrial biology, likely because of the near-universal abundance of water as a solvent, but biological IL formation is known. For example, the tawny crazy ant (*Nylanderia fulva*) detoxifies the alkaloid venom of the fire ant (*Solenopsis invicta*) by combining it with formic acid to form a biologically inert IL [29].

*2.3. Deep Eutectic Solvents as Functional Biological Liquids*

Turning to deep eutectic solvents, a biologically grounded motivation in support of our hypothesis comes from desiccation-tolerant "resurrection" plants. In these cases, naturally occurring DES-like mixtures provide a concrete example in which combinations of simple ionic and hydrogen-bonding components create protective liquid-like phases under conditions in which water loss would normally lead to irreversible denaturation of the cytoplasm.

In plants such as *Selaginella* sp., intracellular mixtures of sugars (e.g., sucrose, trehalose), organic acids, and amino acids form viscous phases during extreme dehydration and are proposed to stabilize proteins, membranes, and cellular ultrastructure over months to years in the absence of bulk water [55]. Upon rehydration, normal metabolic activity resumes.

We note that there is currently no direct proof that liquid natural deep eutectic solvents (NADES; e.g., reviewed in [56]) exist as a distinct macroscopic phase inside living cells. However, substantial circumstantial evidence suggests that they do. Intact plant cells of many species contain all necessary components at sufficient concentrations to form NADES in situ, and in some cases, these components increase markedly under dry conditions, supporting a functional role in dehydration tolerance [57,58]. Similarly, many animals synthesize diverse NADES mixtures as protection against freezing and cold stress [31]. Together, these and other observations (e.g., reviewed in [30,59]) indicate that DES-like liquids may already be integrated into the biochemistry of some life on Earth.

These NADES systems differ fundamentally from the industrial ionic liquids discussed above: they arise from simple metabolites, are biologically regulated, and operate within living cells. In addition, many NADES formulations contain high concentrations of sugars and sugar alcohols, components that are absent in ionic liquids but are central to cellular osmotic balance and macromolecular stabilization.

In conclusion for Sections 2.1–2.3, although most existing studies on the biological compatibility of ILs have been motivated by industrial biocatalysis or biotechnology rather than astrobiology, the studies establish a critical point for the present hypothesis: complex biomolecules can remain folded, soluble, and functional in highly non-aqueous, ionic liquid environments. Additionally, NADES provide evidence that mixtures closely related to DES can function as active biological solvents rather than merely as inert chemical media.

# 3. Implications for Space Exploration

## 3.1. Cosmically Available Building Blocks for ILs and DES

The plausibility of ILs and DES as planetary liquids depends not only on their physical stability, but also on whether their chemical building blocks are available. A wide range of simple anions and cations required for IL and DES formation is known to be abundant throughout the Solar System and in astrophysical environments.

Many anions suitable for abiotic ionic liquid formation occur in planetary settings, including nitrates ($NO_3^-$), hydrogen sulfates ($HSO_4^-$), chlorides ($Cl^-$), perchlorates ($ClO_4^-$), and simple carboxylates such as formate ($HCOO^-$) and acetate ($CH_3COO^-$) (Figure 1 and references therein).

In principle, any positively charged nitrogen-containing organic molecule could serve as the cationic partner (Figure 1), allowing simple anions paired with organic cations to yield stable ionic liquids as plausible outcomes of abiotic chemistry.

Deep eutectic solvents further relax compositional constraints. They can form from mixtures of hydrogen-bond donors and acceptors without full ionization. DES can also readily incorporate metal ions, ammonia ($NH_4^+$), other inorganic and organic cations and anions, mixed with simple polar organics into compound liquids. Such DES can form, e.g., from salts of monovalent cations, such as potassium carbonate ($K_2CO_3$) [14,22,60], or sodium halides (NaCl) [61] mixed with ethylene glycol, glycerol, or carboxylic acids, or divalent cations, like calcium chloride ($CaCl_2$) mixed with sugars and urea [62,63]. DES mixtures containing ammonium chloride mixed with simple organics (e.g., sugar alcohol–sorbitol and urea) have been routinely synthesized [11]. Sugars, alcohols, and carboxylic acids—the simplest components of many DES—are known to exist in planetary environments [64–66] as well as in the interstellar medium and ice grains (reviewed in [67]).

Together, these considerations indicate that the chemical ingredients required to form simple ILs and DES are not exotic, but overlap substantially with known planetary inventories. We now move to individual planetary bodies, expanding on specific ions plausibly available.

## 3.2. Planetary Environments Where IL and DES Could Exist

As part of our hypothesis, we discuss planetary environments where ILs and DES could potentially form and persist. Furthermore, in the Solar System, several environments may provide near-term opportunities for direct investigation. Both archived data and new observations could establish whether ILs or DES exist on Solar System bodies. Existing spacecraft data such as archived Raman and infrared datasets from Mars rovers, asteroid missions, and laboratory analyses of meteorites could be re-examined for spectral features consistent with highly concentrated ionic or eutectic phases, providing a low-cost first pass at constraining their possible presence.

### 3.2.1. Mars and Other Cold Terrestrial Planets

Brines on Mars containing perchlorates, chlorides, and possibly hydrogen sulfates could provide precursors for ILs or DES. Recent work shows that near-surface brines can remain liquid under present-day Martian conditions [68]. Perchlorates have been detected by multiple missions (reviewed in [69]), nitrates are present in sediments [70], and chlorides have been mapped from orbit [71], all representing plausible anionic components for IL or DES formation. As mentioned in Section 3.1, any positively charged nitrogen-containing organic molecule could serve as a cationic partner, even though their presence on Mars remains an open question.

Mars' archived FTIR and Raman instruments could be searched for spectral signatures consistent with IL- or DES-like liquids, guided by laboratory reference spectra. Future landers and subsurface probes could be designed to search for concentrated salt-rich liquids evolving from perchlorate- and chloride-bearing brines, building on existing detections of perchlorates, nitrates, and chlorides and recent demonstrations of liquid brine stability under current conditions [68].

### 3.2.2. Venus Cloud Environments

In the dense haze beneath Venus's clouds, slow accumulation of nitrogen-bearing organics within concentrated sulfuric acid droplets could reach levels sufficient for ionic liquid formation by evaporation of sulfuric acid during long droplet lifetimes. Photochemistry and meteoritic delivery may supply amine-bearing organic cations [72].

In situ Venus's cloud measurements of droplet composition could determine whether nitrogen-bearing organics accumulate in concentrated sulfuric acid aerosols to levels compatible with ionic liquid formation. Spectroscopic constraints on droplet chemistry would again benefit from laboratory FTIR and Raman benchmarks.

3.2.3. Asteroids and Comets

Cometary ices contain chloride, perchlorate, nitrate ions, and nitrogen-bearing organics suitable as IL and DES precursors [73–76]. At comet 67P/Churyumov–Gerasimenko, Rosetta detected phosphorus, halogens, and carboxylic acids, as well as ammonium salts, organic amines, and glycine [77–80]. Urea, alcohol sugars, and other polyols have been detected in the interstellar medium [67,81,82], comets [83], and meteoritic material [64,84,85]. Recent chemical characterization of the material returned from the asteroid Bennu supports the conclusion that organic matter, including the nitrogen-containing organics, is not only common but also can be highly localized and concentrated within the rocky microniches [86].

Both comets and asteroids may host IL- or DES-filled veins below the surface. We emphasize that DES or IL formation on small bodies like comets and asteroids does not necessarily require water sublimation, past hydrothermal activity, or prior dissolution of mixed solid substrates in liquid water. IL and DES can form directly upon mixing of solid components, provided they are sufficiently enriched, and the temperature range allows for the existence of ILs and DES in the liquid form. After formation, ILs and DES can persist protected under the surface for thousands of years between perihelion passages.

It is worth emphasizing that the mixing of ILs (and DES) with water and how dissolution of water affects the chemical and physical properties of the ILs and DES are a major topic to explore. Some preliminary research into this problem (on [Gly][HSO$_4$] IL) shows that such water-IL mixtures may form distinct separate phases [36].

For comets and asteroids, missions capable of sampling protected subsurface material or returning cryogenically preserved samples could test whether IL-like liquids exist at temperatures where water is frozen.

3.2.4. Exoplanets

On rocky exoplanets with thin atmospheres, extreme temperatures, or extensive volatile loss, stable surface water liquid or ice, or atmospheric water vapor may be absent, yet ILs and some DES could remain liquid within pores, fractures, and regolith. Persistence of microscale liquid reservoirs would not require oceans or global hydrological cycles, but only localized chemical concentration and thermal buffering. Such environments could support sustained liquid-phase chemistry even when bulk liquid water is impossible, extending potential habitability to planets previously classified as sterile.

Persistent IL or DES microdroplets, though undetectable remotely, could be sufficient to sustain chemistry and potentially biology. On Earth, viable microbes preserved for millions of years in halite fluid inclusions demonstrate that life can persist in microscopic liquid pockets within solid matrices (e.g., [87,88]). Forgoing reliance on large liquid reservoirs therefore expands the parameter space for habitability to rocky planets, moons, and small bodies long assumed to be barren.

Exoplanets are in a different category for exploration as compared to in situ investigations possible for Solar System planets. On exoplanets, IL- and DES-based liquids are likely to occur primarily as thin films or microscopic droplets within porous rocks or regolith rather than as global surface reservoirs. Such small-scale liquid phases would be effectively invisible to current remote-sensing techniques and would not produce detectable atmospheric or surface spectral signatures. Consequently, the existence of IL- or DES-supported chemistry or biology on exoplanets may remain observationally inaccessible in the near term, even if biologically significant at the microscale. If, however, ILs or DES are present in significant amounts, the effect on surface or atmospheric chemistry may be detectable, as a general property of liquids is that they enable chemistry to happen much faster than between solids. Bulk planetary properties may therefore be changed over time by the presence of trace, 'catalytic' amounts of a liquid. Inferring the presence and identity of that liquid would, however, require deep knowledge of the relevant body's composition and structure, information that requires future ultra–high-angular-resolution imaging or spectroscopy capable of probing planetary surfaces at kilometer or sub-kilometer scales, such as concepts based on the Solar Gravitational Lens Telescope [89] or Starshot-like architectures [90]. Until such capabilities exist, the relevance of ILs and DES to exoplanet habitability will need to be evaluated primarily through laboratory experiments, theoretical modeling, and Solar System analog studies.

## 4. Hypothesized Roles of ILs and DES in Prebiotic Chemistry and Solvent Evolution

We now come to the most exploratory part of our hypothesis, expanding beyond biomolecule compatibility (Section 2) and plausible planetary environments (Section 3). Many questions arise and are topics for future dedicated research (Section 5).

*4.1. Abiotic and Prebiotic Chemistry*

An extreme and speculative part of our IL/DES hypothesis is that prebiotic chemistry could proceed within comets themselves. Comets are already known to host many organic compounds, including carboxylic acids, esters, aldehydes, ketones, alcohols, nitriles, isocyanates, and amides [65,73–76,91]. The organic molecules have been postulated to have originated in the interstellar medium and then incorporated into cometary ices [92–96]. Together with known cations and anions present on comets, such as ammonia, chloride and nitrate, the ingredients for ionic liquid or DES are present [77,79]. In this concept, comets are not merely passive carriers of organics, but chemically active bodies capable of sustaining localized liquid environments even when bulk water remains frozen.

Low temperatures are not, by themselves, prohibitive for complex chemistry or even biology; many terrestrial organisms survive extended deep freezing. One difficulty for water-based scenarios is the physical damage caused by ice formation, which disrupts membranes and macromolecular structures. Ionic liquids and some deep eutectic solvents differ fundamentally in this respect: they solidify without the large volumetric expansion characteristic of water ice, making them intrinsically more compatible with long-term molecular preservation and cryogenic cycling. This property suggests that IL- or DES-rich pockets within cometary material could act as protective phases across repeated extreme temperature cycling.

As comets approach perihelion, ionic liquids—owing to their negligible vapor pressure and protection within pores and fractures—could remain liquid, unlike water ice, which directly sublimates from near-surface layers without reaching a liquid phase. Repeated orbital cycles would then subject these confined liquids to slow heating–cooling and concentration–dilution sequences, progressively enriching dissolved organics and salts. Can complex chemistry and self-organization proceed in these solvents under such conditions? Could cycles of heating and cooling, or of partial drying (or temporary crystallization), drive prebiotic synthesis as they have been hypothesized to do in water [97]? Over millions of years, such micron-scale reservoirs could provide stable settings for sustained reaction networks, allowing chemical systems to advance beyond simple prebiotic synthesis toward increasing molecular complexity—despite potentially dramatic evolution of comet surface and subsurface layers.

Material produced through repeated heating–cooling cycles, concentration–dilution sequences, and confinement within pore-scale liquid phases may go beyond simple processing of raw organics, potentially reflecting extended chemical evolution within a non-aqueous solvent environment and enabling prebiotic pathways that are not thermodynamically favored in water. Chemistry occurring in ILs or DES may therefore offer a possible resolution to the "water problem" in origin-of-life scenarios that require dehydration reactions to form biologically relevant molecules in environments lacking liquid water [96]. Prebiotically generated organic compounds could then be delivered to planetary surfaces by impacts [98]. While many argue that organics are destroyed post-impact (e.g., [99–102]), the counterargument is that a sufficiently large impactor can preserve a substantial fraction of their interior organics, even when surface material is shock-heated or vaporized [103–105]. Preserved material can later be released gradually through fracturing, aqueous alteration, or erosion.

*4.2. Solvent Evolution*

One intriguing, albeit speculative, ramification of IL- or DES-based life is that the solvent itself comes under natural selection and evolutionary pressure. Because living organisms must synthesize their own ILs or DES from available precursors, evolutionary adaptations would drive divergence from the original planetary IL or DES mixture, and from one another. A highly evolved ionic liquid biosphere might therefore contain a wide diversity of ILs or DES compositions that no longer rely on the abiotic, planetary source of the solvent.

A parallel for this chemical diversity exists on Earth. Life uses dozens of different DES-like mixtures as drought and cold protectants, with certain molecules such as trehalose or betaine commonly favored (e.g., [31]). Yet the resulting mixtures differ markedly, even among closely related species like resurrection plants. No two species employ exactly the same component ratios, even when they draw on the same chemical building blocks. In other words, life existing in an IL or DES could adapt both its chemistry and the chemistry of its solvent in parallel, unlike Earth life, which is subject to the unchangeable properties of water.

Speculatively, such a scenario could lead to a complete solvent replacement, a situation where water is replaced by ionic liquids and DES, e.g., as planetary environments dry out. A "water replacement hypothesis" has already been explored for decades, describing in detail how sugars, particularly trehalose, interact with dry membranes and other cellular components to prevent their degradation (e.g., [106–108]).

Extending this transition concept beyond Earth, whether a planet dries out temporarily, permanently, or has always been devoid of water, life could persist by replacing water with ionic liquids or DES and carrying all of its solvent internally. In the temporary case, desiccation is not fatal so long as solvent is retained, allowing organisms to persist much like rare desiccation-tolerant species do on Earth. Over longer timescales, as planetary environments evolve and lose their water permanently, solvent replacement with ionic liquids or DES could sustain survival under radically different conditions. Finally, the possibility of life making its own non-volatile solvent extends habitability even to geologically active worlds that never originated or hosted water-based life. In such a scenario, colonization could occur via panspermia of IL- or DES-based life, as long as the components for biosynthesis of ionic liquids are available, without any reliance on the external planetary reservoirs of ionic liquid.

For background and supportive arguments, collectively, observations from resurrection plants and cold-tolerant animals demonstrate that DES-like mixtures can contribute to maintaining biomolecular structure, membrane integrity, and cellular organization for extended periods in the near-absence of liquid water, with full metabolic activity resuming after rehydration [30,31,56–59,109–111]. Although active metabolism within NADES phases has not been directly demonstrated, the ability of DES-like liquids to preserve complex biochemical systems in a recoverable state establishes that such liquids can sustain the physical and chemical conditions required for biological organization. This provides biological grounding for the hypothesis that DES- and IL-type liquids could play functional roles in non-aqueous biochemistry on other worlds by supporting solvent-mediated chemistry.

## 5. Strategies to Substantiate the IL and DES Hypothesis

Our hypothesis is that ionic liquids and deep eutectic solvents may act as planetary liquids where liquid water is absent, that they might be solvents for non-Earth-like life, and, more speculatively, that they may play a role in prebiotic chemistry on asteroids and comets, and that life elsewhere might purposefully evolve ILs and DES as solvents. Advancing and substantiating this hypothesis will involve laboratory chemistry, computational chemistry, planetary science, and observational strategies.

*5.1. Planetary IL and DES Formation, Stability, and Observational Signatures*

A first class of laboratory studies should establish whether simple ionic liquids and deep eutectic solvents can form abiotically from realistic planetary precursor mixtures. Such experiments should identify efficient formation pathways from cosmically abundant ions and small organic molecules, determine required concentrations and environmental conditions, and characterize long-term chemical stability under vacuum and low-pressure atmospheres. Crystallization behavior over repeated planetary temperature cycles should also be quantified, since solidification would limit the persistence of liquid phases in many environments.

Once formation is established, targeted experiments should measure the fundamental physical and chemical properties of planetary-relevant ILs and DES, including density, viscosity, acidity, melting and crystallization temperatures, decomposition temperatures, and vapor pressure. Many of these properties are missing or are incomplete in the literature (see Table 1). Vapor pressure is particularly challenging to measure for ultra-low-volatility liquids and will require specialized techniques (e.g., [4,112]). Additional experiments should quantify how ionic composition and trace components control and modulate viscosity, polarity, dielectric properties, and reaction kinetics, and should determine the environmental boundaries beyond which DES microdroplets cease to persist as stable liquids, in contrast to ionic liquids.

A separate class of experiments should address tolerance to ultraviolet radiation and high-energy particles. Because many candidate planetary environments involve thin atmospheres or airless bodies, ILs and DES should be exposed to ultraviolet radiation, solar-wind analogs, and energetic particle fluxes to measure degradation rates, radiolytic chemistry, and changes in physical properties under realistic planetary conditions.

Finally, systematic spectroscopic characterization of planetary-relevant ILs and DES using FTIR and Raman scattering is required to build reference libraries of vibrational signatures under controlled temperature and pressure conditions, enabling interpretation of spacecraft, laboratory, and remote-sensing datasets.

*5.2. Biomolecule Stability and Functionality*

Beyond basic physical stability, laboratory work should directly test whether key biological and prebiotic processes can occur in these liquids. Priority targets include: formation and stability of lipid vesicles and other compartment-forming structures; template-directed polymerization and strand separation of nucleic acids; DNA or RNA replication chemistry in low-water or anhydrous regimes; and persistence of catalytic activity in enzyme or ribozyme analog systems. The evidence suggests that ILs could be remarkably compatible with Earth's water-based biochemistry, but the true extent of that compatibility is still unknown.

We specifically identify the need for testing whether self-assembling membranes can form in ionic liquids, because it is a topic yet to be explored in ILs and DES. ILs themselves sometimes do form thin films and membrane-like assemblies that selectively retain certain solutes or are used for ion exchange [113–116]. The formation of supramolecular compartments is considered a prerequisite for life's origins, because such precursors to cell membranes are key for maintaining life's physiology.

*5.3. Computational Chemistry*

Computational chemistry can complement laboratory measurements by predicting key thermodynamic and transport properties of candidate IL and DES compositions prior to synthesis. Ab initio and molecular dynamics simulations can estimate vapor pressure, phase behavior, dielectric properties, solvation structure, and reaction energetics across temperature and pressure ranges that are difficult to reproduce experimentally. Such modeling can narrow the compositional space, guide experimental design, and identify regimes in which non-aqueous liquid phases are most likely to remain stable under planetary conditions.

Beyond physical properties of ILs and DES, computational chemistry can play a key role. Ab initio electronic structure calculations can quantify hydrogen-bonding networks, ion pairing, and reaction energetics relevant to dehydration chemistry, while molecular dynamics simulations can probe peptide folding stability, lipid headgroup–solvent interactions, nucleic acid conformational behavior, and mesoscale organization under non-aqueous conditions. Together, these methods can potentially aid in mapping solvent-dependent free-energy landscapes and identifying thermodynamically plausible regimes for biomolecular stability and prebiotic reaction pathways in planetary IL and DES systems.

*5.4. Modeling IL and DES Microniches and Chemical Survival on Comets and Small Bodies*

An extreme and speculative part of the IL/DES solvent hypothesis is that comets and small bodies could host ILs and DES as tiny reservoirs for prebiotic chemistry. There could be long-lived microniches within pores, fractures, or grain boundaries of rocky or icy material for ionic liquids or, more transiently, deep eutectic solvents to persist where liquid water cannot.

Worth serious exploration are the dramatic changes in comets, e.g., near-surface layers being shed and lost, with lower layers turning into new near-surface layers, etc. Over time—during perihelion passages, long-term orbital evolution, and possibly even diurnal cycles—cometary environmental changes may dynamically modify ILs and DES themselves, altering their composition, concentrating reactants, and exposing these transient solvent phases to high-energy particles in ways analogous to wet–dry cycles that enhance prebiotic chemistry.

Numerical thermal models incorporating realistic orbital histories, internal heat transport, and radiative cooling can be used to identify regions in which ILs or DES could remain liquid while surrounding water ice remains solid or is lost to space. Coupled porosity and diffusion models can then estimate the stability of micron-scale liquid pockets, including susceptibility to evaporation, permeation, or crystallization under low-pressure conditions.

Radiation-transport modeling can further quantify the degree of shielding provided by regolith or ice layers against ultraviolet radiation, solar-wind particles, and galactic cosmic rays, constraining survival times for complex organic molecules and reactive intermediates within such liquid pockets.

## 6. Outlook

The possibility that life might operate in liquids other than water challenges one of the deepest assumptions underlying modern astrobiology. Ionic liquids and deep eutectic solvents, long studied in chemistry and materials science, offer a concrete and physically informed way to extend the concept of habitability into environments previously regarded as irretrievably sterile. Their ability to persist as liquids under extreme temperatures and near-vacuum conditions, to exist in microscopic refuges within rocks and ices, and to coexist with complex organic chemistry invites a rethinking of where and how life-supporting chemistry might occur. Whether or not ILs or DES ultimately prove capable of sustaining living systems, exploring their planetary roles broadens the experimental and conceptual landscape of the search for life. By shifting attention from oceans to micron-scale liquid niches, and from Earth-like conditions to chemically diverse worlds, this hypothesis opens new directions for laboratory research, space mission design, and the interpretation of planetary data. In doing so, it reframes planetary habitability not as a narrow set of environmental criteria but as a spectrum of chemical possibilities still largely unexplored.

**Supplementary Materials:** The following supporting information can be downloaded at https://www.mdpi.com/article/doi/s1, IL Protein Compatibility Database (SI version).xlsx.

**Author Contributions:** Conceptualization: J.J.P., S.S.; investigation: J.J.P., S.S., W.B., I.I., R.A., J.J., P.S.; writing—original draft preparation: J.J.P., S.S.; writing—review and editing: J.J.P., S.S., W.B., I.I., R.A.; funding acquisition: S.S.; All authors have read and agreed to the published version of the manuscript.

**Funding:** This work was partially funded by the NOMIS Foundation Award 1.891 and the Volkswagen Foundation Grant 9E126.

**Data Availability Statement:** All data are included in the manuscript and the Supplementary Materials.

**Acknowledgments:** We thank Maxwell D. Seager for useful discussions. We thank four reviewers for their comments, which substantially improved the article.

**Conflicts of Interest:** The authors declare no conflicts of interest.